\newcommand{\beq} {\begin{eqnarray}}
\newcommand{\eeq} {\end{eqnarray}}
\newcommand{\vev}[1]{ \left\langle {#1} \right\rangle }
\def\unit{\relax{\rm 1\kern-.26em I}}

\documentclass[12pt, preprint,preprintnumbers,amsfonts,aps,nofootinbib]{revtex4}
\usepackage{graphicx}
\begin{document}

\title{Building a Better mSUGRA: WIMP Dark Matter Without Flavor Violation}

\author{Nathaniel J. Craig}
\email{ncraig@stanford.edu}
\affiliation{Department of Physics, Stanford University, Stanford, CA 94305-4060} 

\author{Daniel Green}
\email{drgreen@stanford.edu}
\affiliation{SLAC and Department of Physics, Stanford University, Stanford, CA 94305-4060}

\preprint{SU-ITP-09/27}

\begin{abstract}
The appearance of a natural dark matter candidate, the neutralino, is among the principal successes of minimal supergravity (mSUGRA) and its descendents. In lieu of a suitable ultraviolet completion, however, theories of gravity-mediated supersymmetry breaking such as mSUGRA suffer from arbitrary degrees of flavor violation. Though theories of gauge-mediated supersymmetry breaking are free from such prohibitive flavor violation, they typically lack natural neutralino dark matter candidates. Yet this conventional dichotomy breaks down when the hidden sector is strongly coupled; in models of gauge-mediated supersymmetry breaking, the neutralino may be the lightest supersymmetric particle (LSP) if the fields of the hidden sector possess large anomalous dimensions. In fact, general models of so-called ``sequestered'' gauge mediation possess the full richness of neutralino dark matter found in mSUGRA without corresponding flavor problems. Here we explore generalized models of sequestered gauge mediation and the rich variety of neutralino dark matter they exhibit. 
\end{abstract}

\maketitle
\section{Introduction}

Weak scale supersymmetry (SUSY) is among the most compelling candidates for physics beyond the Standard Model. Among other appealing characteristics, a supersymmetric Standard Model may naturally give rise to weakly-interacting dark matter, provided suitable parities to guarantee the stability of the lightest supersymmetric particle (LSP). The precise nature of the LSP depends sensitively on the way in which supersymmetry breaking is communicated to the fields of the Standard Model. 

Conventional gravity-mediated supersymmetry breaking in the Minimal Supersymmetric Standard Model (MSSM) often provides a suitable dark matter candidate in the form of a neutralino LSP but --  in lieu of a satisfactory ultraviolet completion -- suffers from arbitrary amounts of flavor violation at variance with experimental bounds. Prospects for satisfactory low-energy phenomenology are much more appealing in the case of gauge mediation \cite{Dine:1981za,Dimopoulos:1981au, Nappi:1982hm,AlvarezGaume:1981wy,Dine:1995ag}. Gauge-mediated supersymmetry breaking is free from flavor problems, thanks to the flavor blindness of gauge interactions. However, the flavor-blindness of gauge-mediated supersymmetry breaking has long presented a serious obstacle to producing conventional dark matter candidates. The gravitino mass arises from gravity-mediated effects and thus is set by the Planck scale, $m_{3/2} \sim \frac{F}{M_{Pl}}.$ In contrast, the scale of the remaining sparticle masses is set by the messenger scale $M$ via $\tilde m \sim \frac{F}{M}.$ Since $M \ll M_{Pl}$ in order to avoid reintroducing problematic flavor-violating effects, it is generically the case that $m_{3/2} \ll \tilde m,$ making the gravitino the LSP in all conventional scenarios of gauge mediation. This result is not unique to the MSSM; there are generally gravity-mediated contributions of order $m_{3/2}$ to supersymmetry-breaking mass terms of all sectors, including light hidden sectors with nonstandard dark matter candidates. Even in these more exotic theories, the gravitino is frequently lighter than other states that obtain their mass from supersymmetry breaking.

  Unfortunately, the gravitino is a rather poor candidate for dark matter, both in terms of its cosmology and prospects for direct detection.  Consequently, much effort has been made to engineer satisfactory dark matter candidates within gauge mediation \cite{Dimopoulos:1996gy, Nomura:2001ub, Fujii:2002fv, Ibe:2006rc,Feng:2008zza, Feng:2008ya,Dudas:2008eq}, but such models tend to be somewhat baroque and often fall short of producing signals amenable to direct detection \cite{Akerib:2005kh, Angle:2007uj}.

Conventionally, the specifics of the hidden sector have been considered secondary to the means of mediation in determining the spectrum of the MSSM. However, it has been increasingly observed that the detailed dynamics of the hidden sector may alter na\"{i}ve predictions about low-scale physics \cite{Dine:2004dv, Cohen:2006qc}. Among other possibilities, in models with strongly coupled hidden sectors the scale of gravity-mediated contributions to soft masses may be significantly suppressed. In a purely four-dimensional context, such suppression may be obtained through strong coupling of the hidden sector, an effect known as conformal sequestering \cite{Luty:2001jh,Luty:2001zv}. Conformal sequestering has been explored extensively in the context of anomaly mediation, where it is necessary to suppress gravity-mediated contributions when SUSY breaking occurs at a high scale \cite{Randall:1998uk, Giudice:1998xp}.  

Yet there is no reason for sequestering to pertain only to anomaly mediation. Recently, conformal sequestering was extended to gauge-mediated supersymmetry breaking, where it was shown to solve the $\mu / B \mu$ problem provided certain relations among the anomalous dimensions of operators in a strongly-coupled hidden sector \cite{Roy:2007nz,Murayama:2007ge}. In such a scenario, the gravitino is generally made heavier than the remaining sparticle masses; in \cite{Craig:2008vs} it was demonstrated that this conformal sequestering may alter the prediction of gravitino LSP in even the simplest models of gauge mediation. As a result, the attractive features of gauge mediation could give rise to compelling neutralino dark matter. 

In the minimal single-messenger model of sequestered gauge mediation, the prospects for neutralino dark matter are rather limited; the conformal dynamics necessary to solve $\mu / B \mu$ tend to make the stau rather light, and the neutralino is the LSP in only a small region of parameter space satisfying experimental constraints. However, there is no reason to suppose that SUSY is broken by a single field, or that gauge mediation takes place through a single set of messengers. In this paper, we explore the expanded prospects for neutralino dark matter that arise in even the simplest generalizations of minimal gauge mediation in the presence of a strongly-coupled hidden sector. 

These generalized models of sequestered gauge mediation exhibit the full parametric freedom of mSUGRA while preserving the favorable flavor structure of conventional gauge mediation. As with mSUGRA, these models give rise to cold neutralino dark matter with the observed relic abundance over a wide region of parameter space. Depending on the dynamics of the hidden sector, they may also naturally avoid the $\mu / B \mu$ problem and generate satisfactory electroweak symmetry breaking. Beyond their lack of flavor problems, these models are distinguished from mSUGRA by (1) a supersymmetry breaking scale $\sqrt{F} \sim 10^{11}$ GeV, below which the soft SUSY-breaking parameters run with MSSM renormalization group flow; (2) the optionality of gaugino mass unification, which is present in many models but not a necessary feature; and (3) a direct relation between SUSY-breaking soft parameters and the specific microphysical features of the SUSY-breaking hidden sector. 
	
The organization of this paper is as follows: In Sec. \ref{sec:seq} we review sequestering and its application to gauge mediated supersymmetry breaking, particularly with an eye towards neutralino dark matter. In Sec. \ref{sec:models} we consider various generalizations of single-messenger gauge mediation, including multi-messenger models and `general' gauge mediation, as well as alternate solutions to the $\mu / B \mu$ problem. In Sec. \ref{sec:dm} we focus on the detailed phenomenology of neutralino dark matter in sequestered gauge mediation. In particular, we explore the ways in which sequestered gauge mediation may give rise to the full spectrum of neutralino dark matter conventionally exhibited by gravity-mediated SUSY breaking.

\section{Sequestered Gauge Mediation \label{sec:seq}}

The communication of supersymmetry breaking between a SUSY-breaking hidden sector and the fields of the MSSM generally takes place through higher-dimensional local operators suppressed by powers of a mediation scale $M.$ These operators give rise to supersymmetry-breaking soft masses in the MSSM of the form
\beq
\label{operators}
c_{\phi} \int d^4 \theta \frac{S^{\dag} S}{M^2} \phi^{\dag} \phi \qquad  \qquad c_{W} \int d^2 \theta \frac{S}{M} W_{\alpha} W^{\alpha}
\eeq
where $S$ is a gauge singlet that develops a SUSY-breaking F-term.\footnote{Of course, it is also possible that contributions to soft masses arise from couplings with hidden sector fields $q$ that are gauge non-singlets, but in the interest of simplicity -- and without loss of generality -- in this paper we will focus on gauge singlets $S.$}    Here $\phi$ is an MSSM chiral superfield representing any squark or slepton, while $W_{\alpha}$ are superfields of the MSSM gauge multiplets. When supersymmetry is broken by the $F$-term $F_{S}$, these operators generate SUSY-breaking soft masses for the squarks, sleptons and gauginos. Similar operators give rise to the $\mu, B \mu, A,$ and $B$ parameters of the MSSM. In models of gauge mediation, the mediation scale $M$ is related to the mass of the messenger fields, while in gravity-, anomaly-, and gaugino mediation, it is related to the Planck scale. The explicit form of the coefficients $c_\phi, c_W$ depends on the mechanism of mediation. In the case of gravity mediation, such coefficients generically lead to flavor-changing neutral currents (FCNCs) in violation of experimental bounds, while in the case of gauge mediation, the coefficients of (\ref{operators}) are flavor blind due to the family invariance of gauge interactions. 

\subsection{Conformal sequestering}

	Although one typically expects the coefficients  $c_\phi, c_W$ to be $\mathcal{O}(1),$ this conclusion may be significantly modified by strongly coupled hidden sectors.  Strong dynamics may give the operators $S$ and $S^\dag S$ large, positive anomalous dimensions $\gamma_{S}$ and $\gamma_{S^\dag S},$ respectively.\footnote{Properly speaking, for a strong CFT the nonchiral operator $S^\dag S$ has no well-defined meaning. By ``$S^\dag S$'' we are always referring to the nonchiral operator of lowest scaling dimension in the operator product expansion of $S^\dag \times S$ other than the unit operator. For further discussion of this point, see e.g. \cite{ Luty:2004ye, Grinstein:2008qk, Craig:2009rk}.}
	  Under renormalization group (RG) flow, such large anomalous dimensions will rapidly drive the coefficients $c_\phi, c_W$ towards zero. Assuming the hidden sector becomes strongly coupled at a scale $M_*,$ at renormalization scales $E < M_*$ the coefficients $c_\phi$ and $c_W$ are schematically
\beq
c_\phi(E) = \left(\frac{E}{M_*}\right)^{\gamma_{S^\dag S}} c_\phi(M_*) \qquad \qquad c_W(E) = \left( \frac{E}{M_*} \right)^{\gamma_S} c_W(M_*)
\eeq

	As a result, at the scale of SUSY-breaking one finds $c \ll 1$ for the operators coupling SUSY-breaking to the MSSM.  This mechanism for suppressing SUSY-breaking effects is known as conformal sequestering.
	
	In gravity mediation, when $c_{\phi} \sim \mathcal{O}(1)$, the (flavor violating) contribution to the scalar masses is at the same scale as the gravitino mass.  A natural solution to the flavor problem is to make the gravitino mass much smaller than the weak scale and generate the scalar masses at the weak scale by a flavor blind mechanism such as gauge mediation.  In these models the gravitino is inevitably the LSP, and there are no natural neutralino dark matter candidates. Conformal sequestering, however, offers a natural alternative to the gravitino LSP.  By forcing $c_{\phi} \ll 1$, the gravitino mass can be above the weak scale without introducing significant flavor violation. In particular, strong dynamics in the hidden sector may change the relation between the masses of gauginos and the gravitino. At the SUSY-breaking scale $\sqrt{F_S},$ the gaugino, scalar, and gravitino masses are given respectively by
\beq
M_a = \frac{\alpha_a}{4 \pi} \left( \frac{\sqrt{F_S}}{M} \right)^{\gamma_S} \frac{F_S}{M}
 \hspace{1cm} 
 m_{\tilde f}^2 = 2 C_r^{\tilde f} \left( \frac{\alpha_r}{4 \pi} \right)^2 \left( \frac{\sqrt{F_S}}{M} \right)^{\gamma_{S^\dag S}} \frac{F_S^2}{M^2}
\hspace{1cm} 
m_{3/2} = \frac{F_S}{\sqrt{3} M_{Pl}}
\eeq
In the weakly-coupled case where $\gamma_S \simeq 0,$ one finds the familiar result $M_a / m_{3/2} = (\alpha_a/4 \pi) (M_{Pl} / M).$ The smallness of flavor-violating effects from Planck slop requires $M / M_{Pl} \simeq 10^{-4},$ implying that the gravitino is always significantly lighter than the gauginos. In the strongly coupled case where $\gamma_S, \gamma_{S^\dag S} \sim \mathcal{O}(1),$ however, the extra factor $ \left( \frac{\sqrt{F_S}}{M} \right)^{\gamma_S}$ may naturally lower the gaugino masses relative to the gravitino, allowing for neutralino dark matter without the reintroduction of flavor problems.\footnote{It is worth emphasizing that the omnipresent gravity-mediated contribution to the gaugino masses also obtains a suppression $ \left( \frac{\sqrt{F_S}}{M} \right)^{\gamma_S},$ so that flavor problems are not reintroduced when $M_a \lesssim m_{3/2}.$}
	
	We will refer to this alteration of the relationship between gaugino and gravitino masses by gauge mediation of supersymmetry breaking from a strongly-coupled hidden sector as `Sequestered Gauge Mediation'. Sequestered gauge mediation provides a natural mechanism for realizing weak-scale neutralino dark matter in a flavor-blind setting without recourse to baroque model-building, much in the spirit of \cite{Luty:2002ff}. It should be emphasized that the potential for sequestering to produce viable dark matter extends well beyond neutralino dark matter in the MSSM. In particular, it may allow more exotic non-gravitino LSP's other than the neutralino, including light fields in `dark sectors' such as those proposed to explain recent astrophysical anomalies \cite{Morrissey:2009ur}.

\subsection{Scalar sequestering}	
	
	Sequestered gauge mediation is also known to provide a natural solution to the $\mu / B \mu$ problem \cite{Roy:2007nz,Murayama:2007ge}.  By adding direct couplings of the Higgs to the messengers, we generate the operators
\beq
\label{cmu}
c_{B \mu} \int d^4 \theta \frac{S^{\dag} S}{M^2} H_{u} H_{d} \qquad  \qquad c_{\mu} \int d^4 \theta \frac{S^{\dag}}{M} H_{u} H_{d}. 
\eeq
When supersymmetry is broken by the F-term of $S$, we arrive at $B\mu$ and $\mu$ terms for the Higgs sector.  Successful electroweak symmetry breaking requires $\mu^2 \simeq B\mu$ which implies $c_{\mu}^2 \simeq c_{B\mu}$.  In weakly-coupled theories of gauge mediation, both operators are typically generated at one loop, resulting in $c_{\mu}^2 \ll c_{B\mu}$. This is the origin of $\mu / B\mu$ problem in gauge mediation.

The coefficients in (\ref{cmu}) are generated at the scale $M$ where the messengers are integrated out.  Conformal sequestering below this scale may naturally resolve the $\mu / B\mu$ problem.  At a strongly coupled fixed point point, there is no reason for the anomalous dimensions of $S$ and $S^\dag S$ to be related.  If $2 \gamma_{S} < \gamma_{S^{\dag}S}$, then $c_{B\mu}$ will run towards zero more quickly then $c_{\mu}^2$, thus allowing for electro-weak symmetry breaking at low energies despite the hierarchy between $c_{\mu}^2$ and $c_{B\mu}$ at high energies. 

\begin{figure}[t] 
   \centering
   \includegraphics[width=4in]{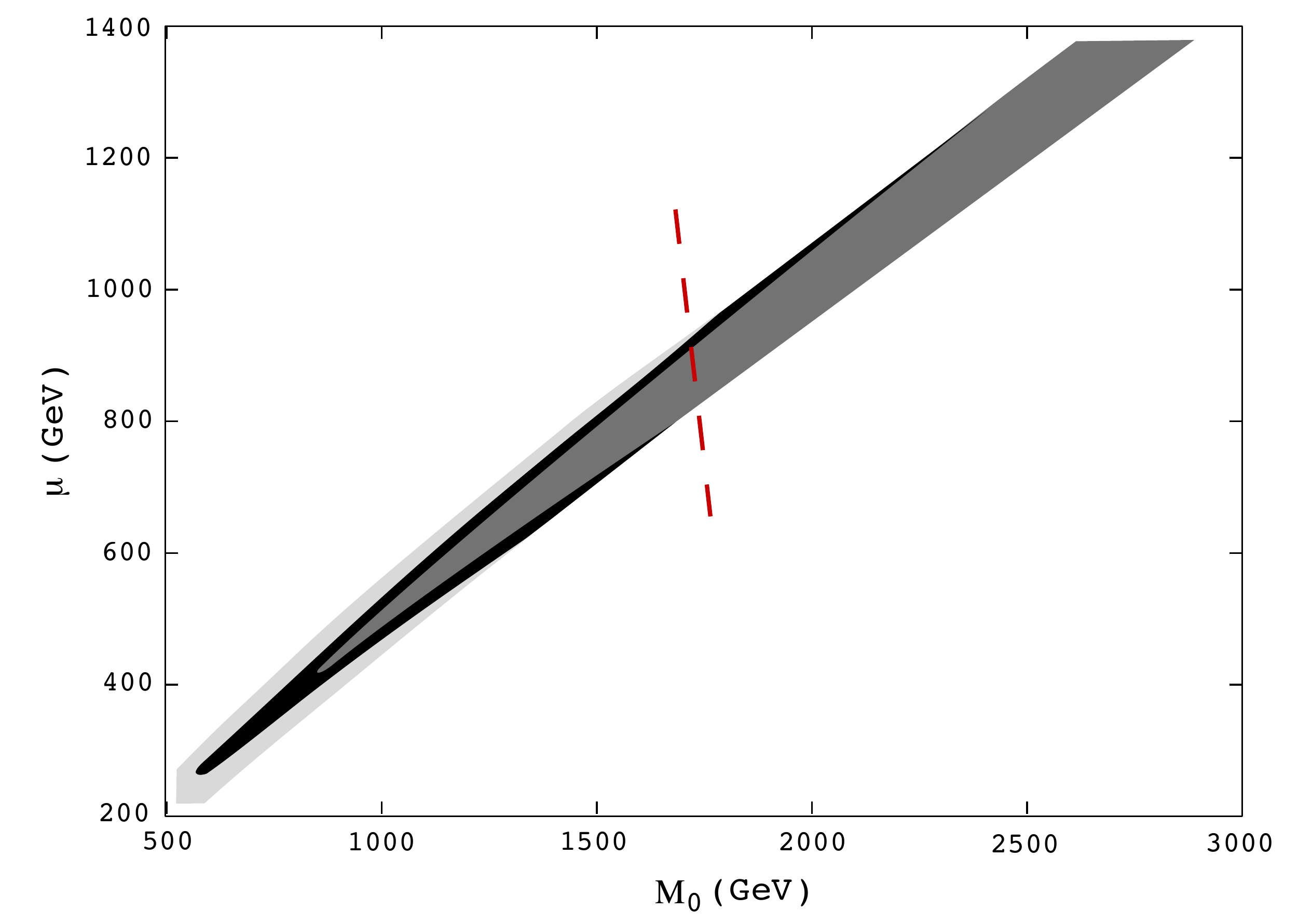} 
   \caption{Parameter space for neutralino dark matter in single-messenger sequestered gauge mediation with scalar sequestering as a function of $\mu$ and unified gaugino mass $M_0 \equiv  (\sqrt{F_S} / M)^{1 + \gamma_S} / 4 \pi.$ Here $\sqrt{F_S} = 2 \times 10^{11}$ GeV, $\tan \beta = 5.0,$ $B \mu \simeq \mu^2,$ and $\tilde m (\sqrt{F_S}) \simeq M_0  / 4 \pi.$ Regions with adequate relic abundance ($0.094 \leq \Omega h^2 \leq 0.129$) are shown in black; regions with inadequate relic abundance ($\Omega h^2 < 0.094$) are shown in light grey; regions with excessive relic abundance ($\Omega h^2 > 0.129$) are shown in dark grey. The light higgs mass satisfies $m_h \geq 111.4$ GeV to the right of the dotted red line. Regions excluded by LEP bounds, $B \rightarrow X_s \gamma,$ or direct detection are white.}
   \label{fig:min}
\end{figure}

Solving the $\mu$ problem in this way has consequences for the rest of the spectrum.  The relative running between $\mu$ and $B\mu$ must also appear between the gaugino masses and the scalar masses.  Specifically, this will drive the scalar masses to small values at the scale $\sqrt{F_S}$.  In single messenger models, at the scale $M$ we have $c_{W}^2 \simeq c_{\phi}$.  Therefore, at the SUSY-breaking scale $\sqrt{F_S}$ we would have $M_a^2(\sqrt{F_S}) = \alpha_a^2 M_0^2 \gtrsim 16 \pi^2 \tilde m^2 (\sqrt{F_S})$, where $M_0 \equiv (\sqrt{F_S} / M)^{1 + \gamma_S} (\sqrt{F_S}/ 4 \pi)$ is the unified gaugino mass.  This characteristic spectrum arising from a conformally-sequestered solution to the $\mu / B \mu$ problem has come to be known as `scalar sequestering'. In \cite{Craig:2008vs} it was demonstrated that the spectrum of scalar sequestering places a significant constraint on the available parameter space for neutralino dark matter. As shown in Fig. \ref{fig:min}, neutralino dark matter with suitable relic abundance arises in only a narrow region of parameter space for minimal single-messenger models of scalar sequestering.  The focus of this paper is expand the parameter space for neutralino dark matter by considering more general mechanisms of gauge mediation.

\section{Models of Sequestered Gauge Mediation \label{sec:models}}

Although minimal single-messenger models of gauge mediation are often favored for their simplicity, there is no reason to assume the messenger sector is so straightforward. There exist a cornucopia of models in which the numbers, representations, and couplings of messenger and hidden sector fields vary (for a review of early models, see \cite{Giudice:1998bp}; for recent efforts to parameterize the cornucopia, see \cite{Meade:2008wd}). Even simple generalizations of the minimal model may result in new degrees of parametric freedom among SUSY-breaking soft masses and markedly different weak-scale spectra. Here we wish to consider a few representative generalizations of gauge mediation for theories in which the hidden sector is strongly coupled. Such generalized models often yield independent scales for gaugino and scalar masses, which may significantly expand the parameter space for neutralino dark matter.

The scales of these theories will be determined by the desire for a non-gravitino LSP, in which case the gravitino mass is larger than the other soft masses at the weak scale. Given $m_{\frac{3}{2}} \sim F_S M_{Pl}^{-1},$ this suggests a high gauge-mediated supersymmetry-breaking scale $\sqrt{F_S} \gtrsim 10^{10-11}$ GeV. Since there is a gravity mediation contribution to $B\mu$ proportional to $\mu m_{\frac{3}{2}},$ preserving the sequestered solution to the $\mu$ problem likewise indicates $\sqrt{F_S} \sim 10^{11}$ GeV.

The exact boundary conditions for soft SUSY-breaking terms at the scale of supersymmetry breaking are fixed by the strong dynamics of the hidden sector. In general, the spectrum of soft masses is of the form
\beq
M_a^2 \sim \mu^2 \sim m_{H_u, H_d}^2 \sim A_{u,d}^2 
\eeq
For general theories of sequestered gauge mediation, the size of  $B \mu$ and the squark and slepton masses $\tilde m$ depends on the relation between $\gamma_S$ and $\gamma_{S^\dag S}.$ In the case of scalar sequestering, $2 \gamma_S < \gamma_{S^\dag S}$ leads to $B \mu \sim \mu^2$ and $\tilde m^2 \ll M_a^2, \mu^2,$ etc. There has been some dispute in the literature regarding the exact relation between $m_{H_u, H_d}^2$ and $\mu^2$ at the scale of supersymmetry-breaking; a general prescription is given in \cite{Craig:2009rk}, and depends on the dynamics of the hidden sector CFT. Given this parametric freedom between $m_{H_u, H_d}^2$ and $\mu^2$ at the scale of SUSY-breaking, we choose to vary $M_0, \mu,$ $B \mu$ at the scale $\sqrt{F_S}$ and $\tan \beta$ at the scale of EWSB ; $|m_{H_u, H_d}^2| \sim \mu^2$ are then chosen to satisfy the conditions for electroweak symmetry breaking.

In order to compute the dark matter relic abundance in these scenarios, we have used the supersymmetric spectrum calculator \texttt{SOFTSUSY} \cite{Allanach:2001kg} and the dark matter software package \texttt{MicrOMEGAs} \cite{Belanger:2001fz, Belanger:2004yn, Belanger:2006is, Belanger:2008sj}. The LSP parameter space is constrained by various experimental and theoretical bounds. Foremost is the WMAP bound on dark matter relic abundance, $0.094 < \Omega_{DM} h^2 < 0.129$ \cite{Dunkley:2008ie}. Though flavor invariance of SUSY-breaking soft terms is one of the principal successes of gauge mediation, constraints from FCNCs may still arise from processes such as the B-meson decay $B \rightarrow X_s \gamma;$ here we restrict ourselves to parameter space satisfying the most recent HFAG global average $\mathcal{B}(B \rightarrow X_s \gamma) = (3.52 \pm 0.23 \pm 0.09) \times 10^{-4}$ \cite{Barberio:2008fa}. The lightness of sleptons in many models of sequestered gauge mediation implies that experimental bounds from LEP2 play a significant role in constraining the dark matter parameter space; we require the low energy spectrum to satisfy the lower bounds on sparticle masses from non-observation at LEP2 \cite{Barate:2003sz, Alcaraz:2007ri}. In particular, we take account of direct search bounds on the lightest neutral Higgs mass, $m_h > 114.4$ GeV, allowing for $\pm 3$ GeV due to theoretical errors among spectrum-calculating software packages. We are likewise interested in satisfying rudimentary constraints on the stability of the Higgs scalar potential. Requiring that the Higgs scalar potential possesses nontrivial extrema (i.e., $\langle h_u^0 \rangle, \langle h_d^0 \rangle \neq 0$) entails $(B \mu)^2 > (m_{H_u}^2 + \mu^2) (m_{H_d}^2 + \mu^2),$ while ensuring that the scalar potential possesses a stable minimum leads to $m_{H_u}^2 + m_{H_d}^2 + 2 \mu^2 > 2 | B\mu |.$ Although experimental and theoretical constraints on the low-energy spectrum are numerous, they may be satisfied over a wide range of parameter space in generalized models of sequestered gauge mediation.

\subsection{Sequestered Gauge Mediation without Scalar Sequestering}

To date, most models of sequestered gauge mediation have focused on solutions to the $\mu / B \mu$ problem via strong dynamics. This approach requires a specific relation between anomalous dimensions of hidden-sector operators, $2 \gamma_S < \gamma_{S^\dagger S},$ and results in the distinctive spectrum of ``scalar sequestering'' in which soft masses for squarks and sleptons are suppressed relative to those for gauginos. As a result, the low-energy spectrum of these theories often possess stau or sneutrino LSP, and the region of parameter space yielding neutralino LSP is highly constrained. 

 However, from the perspective of dark matter, the particular relation between $\gamma_S$ and $\gamma_{S^\dagger S}$ is unimportant. Indeed, it may be more ``generic'' for strongly-coupled hidden sectors {\it not} to obey the relations required by scalar sequestering, since to date there are no tractable 4d $\mathcal{N} = 1$ CFTs with $2 \gamma_S < \gamma_{S^\dagger S}.$  The principal effect of conformal sequestering in these scenarios is to raise the gravitino mass relative to the masses of the other MSSM sparticles. There are certainly many weakly coupled solutions to the $\mu / B \mu$ problem of gauge mediation \cite{Dvali:1996cu,Dimopoulos:1997je, Langacker:1999hs,Hall:2002up, Giudice:2007ca,Liu:2008pa, Green:2009mx}, making it reasonable consider theories wherein the solution to $\mu / B \mu$ arises from dynamics unrelated to sequestering. 
 
\begin{figure}[t] 
   \centering
   \includegraphics[width=4in]{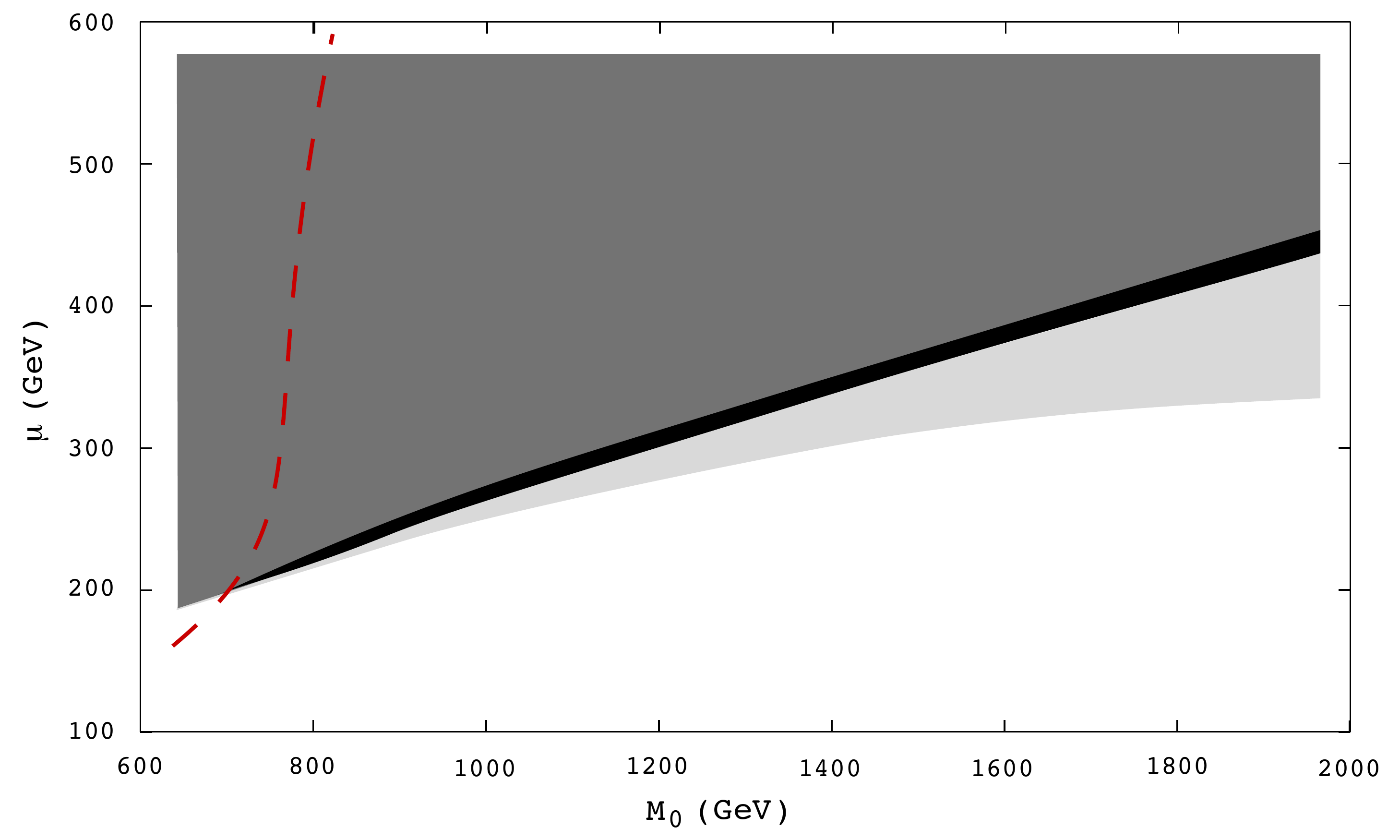} 
   \caption{Parameter space for neutralino dark matter in single-messenger sequestered gauge mediation {\it without} scalar sequestering as a function of unified gaugino mass $M_0$ and $\mu$, assuming some other solution for the $\mu / B \mu$ problem. Here $\sqrt{F_S} = 10^{11}$ GeV, $\tan \beta = 10.0,$ $B \mu \simeq \mu^2,$ and $\gamma_{S^\dag S} = 2 \gamma_S.$ Color conventions are as in Fig. \ref{fig:min}}
   \label{fig:nbm}
\end{figure}

Generally speaking, $\gamma_{S}$ and $\gamma_{S^\dagger S}$ are independent; if strongly coupled dynamics is no longer responsible for solving the $\mu$ problem, there is no reason to assume any relative relation. However, in most tractable 4d CFTs, the anomalous dimensions obey $2 \gamma_S \simeq \gamma_{S^\dagger S}$ to good approximation. This is certainly true of large-N CFTs, where $\gamma_{S^\dag S} = 2 \gamma_S + \mathcal{O}(1/N).$ It is likewise true of the CFT duals of 5d Randall-Sundrum (RS) models for essentially the same reason; they are also large-N theories. Then perhaps the most generic relationship to expect is $2 \gamma_S = \gamma_{S^\dagger S},$ in which case operators quadratic and linear in $S$ receive the same degree of sequestering. Consequently, high-scale squark and slepton masses are of the order of the gaugino masses at the SUSY-breaking scale, and MSSM running down to the weak scale results in generic neutralino dark matter for a wide range of parameters. A representative slice of parameter space is shown in Fig. \ref{fig:nbm}.

\subsection{Multi-messenger Models with Scalar Sequestering  \label{sec:multimess}}

That being said, a conformally sequestered hidden sector with $2 \gamma_S \lesssim \gamma_{S^\dagger S}$ would simultaneously provide both a compelling solution to $\mu / B \mu$ and prospects for neutralino dark matter in gauge mediation. While the parameter space for neutralino dark matter is highly constrained in single-messenger sequestered gauge mediation, these constraints are relaxed by simple generalizations to more complicated messenger sectors. 

Perhaps the most straightforward generalization of conventional single-messenger gauge mediation is to the case of multiple messengers and SUSY-breaking gauge singlets. To this end, consider a theory with $n$ $10 + \overline{10}$ messenger pairs $\Phi_i, \bar \Phi_i$ ($i = 1, ... , n$), as explored in \cite{Dimopoulos:1996ig}. These messengers are coupled to $n \times n$ singlet superfields $S_{ij}$ obtaining SUSY-breaking F-terms $F_{ij}.$ The messenger mass matrix is contained in the superpotential
\beq
W = \bar \Phi_i M_{\bar{i} j} \Phi_j \hspace{1cm} i,j = 1, ..., n
\eeq
and a SUSY-breaking term in the scalar potential
\beq
V = \bar \Phi_i F_{\bar{i} j} \Phi_j + h.c.
\eeq 
For simplicity, we assume $F$ and $M$ are generated by a sector neutral under GUT interactions, so that at the GUT scale $F$ and $M$ are the same for both doublets and triplets. Assuming $M_{ij}$ and $F_{ij}$ are general $n \times n$ matrices, the resultant SUSY-breaking soft masses are parametrically different from the conventional single-messenger case.  In order to generate $\mu$ and $B \mu$, we will also add direct couplings of the messengers to the Higgs through the superpotential terms
\beq
W \supset \lambda_{ij} H_{u} \Phi_{i} \Phi_{j} + \lambda_{\bar{i}\bar{j}} H_{d} \bar{\Phi}_{\bar i} \bar{\Phi_{\bar j}}.
\eeq
If we demand the theory is invariant under $\bar{\Phi}^{\dag} \leftrightarrow \Phi$ (i.e. messenger parity), then $M_{i \bar{j}}^{\dag} = M_{j \bar{i}}$, $F_{i \bar{j}}^{\dag} = F_{j \bar{i}}$ and $\lambda_{ij}^* = \lambda_{\bar{i} \bar{j}}$.

Naturally, the anomalous dimensions $\gamma_{S_{ij}}$ (and, likewise, $\gamma_{S_{ij}^\dagger S_{ij}}$) of the SUSY-breaking singlets $S_{ij}$ need not be identical; it is quite possible that operators corresponding to different $S_{ij}$ receive differing degrees of sequestering in the conformal regime.  In such cases, it may be unnatural to expect the gaugino and scalar masses to be at the weak scale.  Therefore it will be useful to consider cases where there the anomalous dimensions are identical.  For example, if the F-terms are generated by a single field S that couples differently to messengers through $W \supset \lambda_{\bar{i} j} \bar{\Phi}_{\bar{i}} S \Phi_{j}$, then all the anomalous dimensions satisfy $\gamma_{S_{ij}}= \gamma_{S}$ and $\gamma_{S_{ij}^\dagger S_{ij}} = \gamma_{S^\dagger S}$.  In such an example, we assume that the mass matrix $M_{\bar{i} j}$ is {\it not} generated by a vacuum expectation value for S, but is a tree level mass term.

Assuming the hidden sector enters a conformal regime at a scale $M_* \sim M_{ij}$ and leaves it at a scale $\Lambda_* \sim \sqrt{F_{ij}},$ effective operators quadratic in $S$ receive a schematic suppression $(\Lambda_* / M_*)^{\gamma_{S_{ik}^\dagger S_{jl}}},$ while those linear in $S$ receive a suppression $(\Lambda_* / M_*)^{\gamma_{S_{ij}}}.$

After suitable redefinition of the messenger superfields, we can choose $M_{ij}$ to be diagonal with real and positive eigenvalues $M_i$ ($i = 1, ..., n$). The gaugino masses $M_a$ at the SUSY-breaking scale $\sqrt{F_S}$ are then schematically
\beq
M_a = 3 \frac{\alpha_a}{4 \pi} \sum_{i = 1}^n \left(\frac{\Lambda_*}{M_*} \right)^{\gamma_{S_{ii}}} \frac{F_{ii}}{M_i} \equiv \frac{\alpha_a}{4 \pi} \Lambda_G
\eeq
 Here and henceforth, we assume the entries of $F$ are less than those of $M^2,$ a reasonable approximation. 

The two-loop contributions to squark and slepton masses are somewhat more complicated; in the limit where $M$ is proportional to the identity, $M = M_0 \unit,$ they obtain the form 
\beq
m_{\tilde f}^2 = 6  \sum_{r = 1}^3 C_r^{\tilde f} \left( \frac{\alpha_r}{4 \pi} \right)^2 \sum_{i,j=1}^{n} \left(\frac{\Lambda_*}{M_*} \right)^{\gamma_{S_{ij}^\dagger S_{ij}}}\frac{|F_{ij}|^2}{M_0^2} \equiv 2 \sum_{r = 1}^3 C_r^{\tilde f} \left( \frac{\alpha_r}{4 \pi} \right)^2 \Lambda_S^2
\eeq
In general, $\Lambda_G$ and $\Lambda_S$ are independent, and may be treated as distinct parameters; hence the gaugino masses $M_a$ and squark \& slepton masses $m_{\tilde f}$ are parametrically separate.  It is also possible that squark and slepton masses-squared may be generated at one loop via D-term contributions, which may be prohibitively large. Such contributions vanish in the case of messenger parity, and so may be neglected.  

Due to the direct coupling to the Higgs, both $\mu$ and $B \mu$ are generated at one loop.  The $\mu$ term takes the form
\beq
\mu = \frac{3}{16 \pi^2} \sum_{i,j,k=1}^{n} \lambda_{i k} \lambda_{\bar j \bar k} \left(\frac{\Lambda_*}{M_*} \right)^{\gamma_{S_{ij}}} \frac{F_{ij}}{M_i},
\eeq
while the $B\mu$ term takes the form
\beq
B\mu = \frac{3}{16 \pi^2} \sum_{i,j,k,l=1}^{n} \left( \lambda_{i k} \lambda_{\bar j \bar l} \left(\frac{\Lambda_*}{M_*} \right)^{\gamma_{S_{ik}^\dagger S_{j l}}}\frac{F_{ij} \bar{F}_{kl}}{M_0^2} +\lambda_{i k} \lambda_{\bar j \bar k} \left(\frac{\Lambda_*}{M_*} \right)^{\gamma_{S_{il}^\dagger S_{j l}}}\frac{F_{il} \bar{F}_{jl}}{M_0^2} \right).
\eeq
The direct coupling to the Higgs also generates $A$ terms.  For example, the operators $\int d^4\theta \frac{S}{M} H^{\dag}_{u}H_{u}$ will generate $A$ terms for all the fields coupling to the up-type Higgs.  This $A$-term is given by
\beq
A_{u} = \frac{3}{16 \pi^2} \sum_{i,j,k=1}^{n} \lambda_{i k} \lambda^*_{j k} \left(\frac{\Lambda_*}{M_*} \right)^{\gamma_{S_{ij}}} \frac{F_{ij}}{M_i}.
\eeq
For the $A$ term for down sector, one simply replaces $\lambda_{ij}$ with $\lambda_{\bar{i} \bar{j}}$.  If we impose messenger parity, then it follows that $\lambda_{\bar{i} \bar{j}} = \lambda^*_{ij}$ and that $A_{u} = A_{d} = \mu$.

The expression for $\mu$ and $B\mu$ are sightly complicated, so we will consider two cases that illustrate the different possibilities.  First of all, consider the case where $\lambda_{ij} = \lambda_{\bar{i} \bar{j}} = \delta_{ij}$.  In this case, we get the following relation:
\beq
\frac{B\mu}{\mu^2} = 16\pi^2 \frac{\Lambda_{S}^2}{\Lambda_{G}^2}.
\eeq
The important thing to note is that we now have worsened the $\mu$ problem by exactly the same parametric separation as we achieved between the gaugino and scalar masses.  As a result, this would do nothing to change our boundary conditions at $\Lambda_{*}$.  The second case we will consider is where the only non-zero Higgs couplings are $\lambda_{11} = \lambda_{\bar1 \bar 1} = 1$.  In this case,
\beq
\mu =\frac{3}{16 \pi^2} \left(\frac{\Lambda_*}{M_*} \right)^{\gamma_{S_{11}}} \frac{F_{11}}{M_0},
\eeq
and
\beq
B\mu = \frac{3}{16 \pi^2} \left( \left(\frac{\Lambda_*}{M_*} \right)^{\gamma_{S_{11}^\dagger S_{11}}}\frac{F_{11} \bar{F}_{11}}{M_0^2} +\sum_{l=1}^{n} \left(\frac{\Lambda_*}{M_*} \right)^{\gamma_{S_{1l}^\dagger S_{1 l}}}\frac{F_{1l} \bar{F}_{1l}}{M_0^2}   \right).
\eeq
While the $B\mu$ term still contains a sum over F terms, one should note that many contributions to $m_{\tilde f}^2$ do not appear in $B\mu$.  Therefore, this model will allow for parametric separation of the gaugino and scalar masses, without an associated separation of $\mu$ and $B\mu$.

\begin{figure}[t] 
   \centering
   \includegraphics[width=4in]{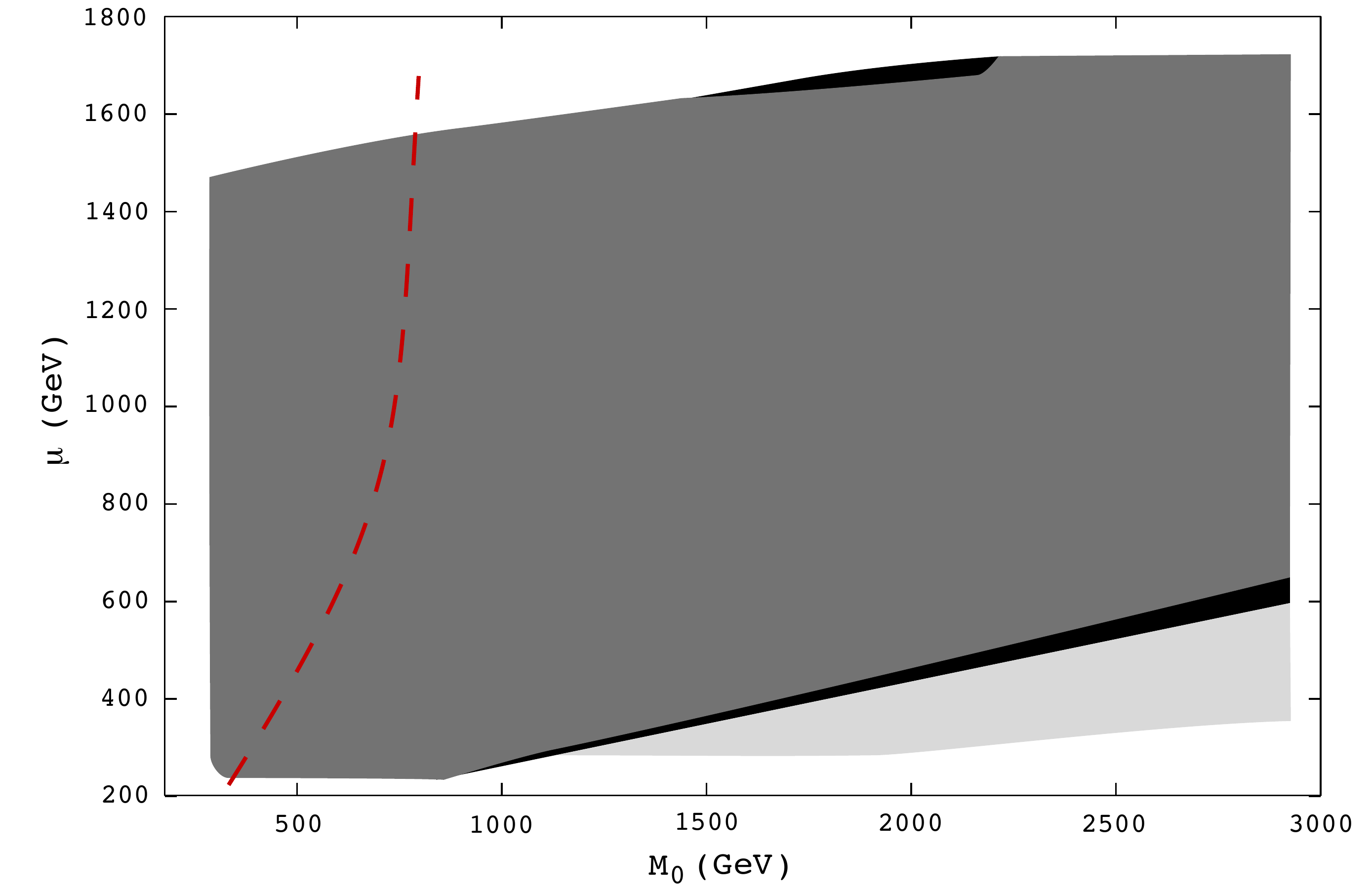} 
   \caption{Parameter space for neutralino dark matter in multi-messenger sequestered gauge mediation with scalar sequestering as a function of $\mu$ and unified gaugino mass $M_0 \equiv 4 \pi \Lambda_G. $ The only non-zero couplings between the Higgses and messenger fields in this model are $\lambda_{11} = \lambda_{\bar 1 \bar 1},$ so that the $\mu / B \mu$ problem may be solved without a hierarchy between gaugino and scalar masses. Here $\sqrt{F_S} = 10^{11}$ GeV, $\tan \beta = 10.0,$ $B \mu \simeq \mu^2,$ and $\tilde m = 1$ TeV. Color conventions are as in Fig. \ref{fig:min}}
   \label{fig:mm1}
\end{figure}

The important feature of these models is that, when the Higgs couples preferentially to one set of messengers, the scale of the gaugino masses $\Lambda_{G}$ and the scale of the scalar masses $\Lambda_{S}$ can vary independently.  Whereas solving the $\mu$ problem in the case of single-messenger mediation led to the high-scale constraint $m_{\tilde f}(\sqrt{F_S}) \sim M_0(\sqrt{F_S}) / 4 \pi$ -- a significant restriction on the parameter space of neutralino dark matter -- in the multiple-messenger case it allows gaugino and slepton masses to vary independently while retaining a conformally-sequestered solution to the $\mu$ problem. This significantly expands the regime of neutralino dark matter, as seen in Fig. \ref{fig:mm1}.

\subsection{General Gauge Mediation with Scalar Sequestering \label{sec:ggm}}
 
An even more general parameterization of gauge-mediated supersymmetry breaking has been made in  \cite{Meade:2008wd}, which explores the most general possible set of parameters generated by direct or indirect gauge mediation, whether weakly or strongly coupled. The result is a six-parameter theory with independent gaugino and sfermion masses and a high degree of parametric freedom. 

For aesthetic simplicity, here we will focus on the weakly-coupled realizations of general gauge mediation elaborated in \cite{Carpenter:2008wi}, in which many of the parameters of \cite{Meade:2008wd} are generated without recourse to further incalculable strongly-coupled physics. Using such weakly-coupled models, it is not terribly difficult to obtain gauge-mediated theories with non-universal gaugino masses. Perhaps the simplest possible example involves a single set of $5 + \bar 5$ messengers and a hidden sector singlet $S$, with superpotential
 \beq
 W = S \left( \lambda_q q \tilde q + \lambda_l l \tilde l \right) + M (q \tilde q + l \tilde l)
 \eeq
which exhibits a messenger-parity symmetry. For simplicity, we consider the case in which the messengers share a common mass $M$ that dominates over the contribution from $\vev{S};$ it is straightforward to generalize to more complicated cases. We assume the singlet $S$ obtains a SUSY-breaking F-terms $F,$ and that the hidden sector enters a conformal regime at a scale $M_* \lesssim M$ and leaves it at a scale $\Lambda_* \gtrsim \sqrt{F_S}.$   At the scale of SUSY breaking, the spectrum of gaugino masses is described by two parameters,
\beq
\Lambda_q \equiv \lambda_q  \frac{F}{M} \left( \frac{\Lambda_*}{M_*} \right)^{\gamma_{S}} \hspace{5mm} \Lambda_l \equiv \lambda_l  \frac{F}{M} \left( \frac{\Lambda_*}{M_*} \right)^{\gamma_{S}} 
\eeq
In terms of these two parameters, the gaugino masses are given by
\beq
M_1 = \frac{\alpha_1}{4 \pi} \left[ \frac{2}{3} \Lambda_q +  \Lambda_l \right] \hspace{5mm} M_2 = \frac{\alpha_2}{4 \pi} \Lambda_l \nonumber \hspace{5mm} M_3 = \frac{\alpha_3}{4 \pi} \Lambda_q 
\eeq
In the original, weakly-coupled version of this model outlined in \cite{Carpenter:2008wi}, the squark and slepton masses-squared depend simply on $\Lambda_q^2, \Lambda_l^2.$ However, in the strongly-coupled case these soft masses are generated by operators quadratic in $S,$ and so depend on the anomalous dimensions $\gamma_{S^\dagger S}.$ Thus the squark and slepton masses depend on two related parameters,
\beq
\tilde \Lambda_q^2 \equiv \left| \frac{\lambda_q F}{M} \right|^{2} \left( \frac{\Lambda_*}{M_*} \right)^{\gamma_{S^\dagger S}} \\
\tilde \Lambda_l^2 \equiv \left| \frac{\lambda_l F}{M} \right|^{2} \left( \frac{\Lambda_*}{M_*} \right)^{\gamma_{S^\dagger S}}
\eeq

In general, for example, $\tilde \Lambda_q^2 \neq \Lambda_q^2.$ In terms of these parameters, the squark and slepton masses are given by
\beq
m_{\tilde f}^2 = 2 \left[ C_1^{\tilde f} \left( \frac{\alpha_1}{4 \pi} \right)^2  \left(\frac{2}{3} \tilde \Lambda_q^2 + \tilde \Lambda_l^2 \right) + C_2^{\tilde f} \left( \frac{\alpha_2}{4 \pi} \right)^2 \tilde \Lambda_l^2 + C_3^{\tilde f} \left( \frac{\alpha_3}{4 \pi} \right)^2 \tilde \Lambda_q^2 \right]
\eeq

\begin{figure}[t] 
   \centering
   \includegraphics[width=4in]{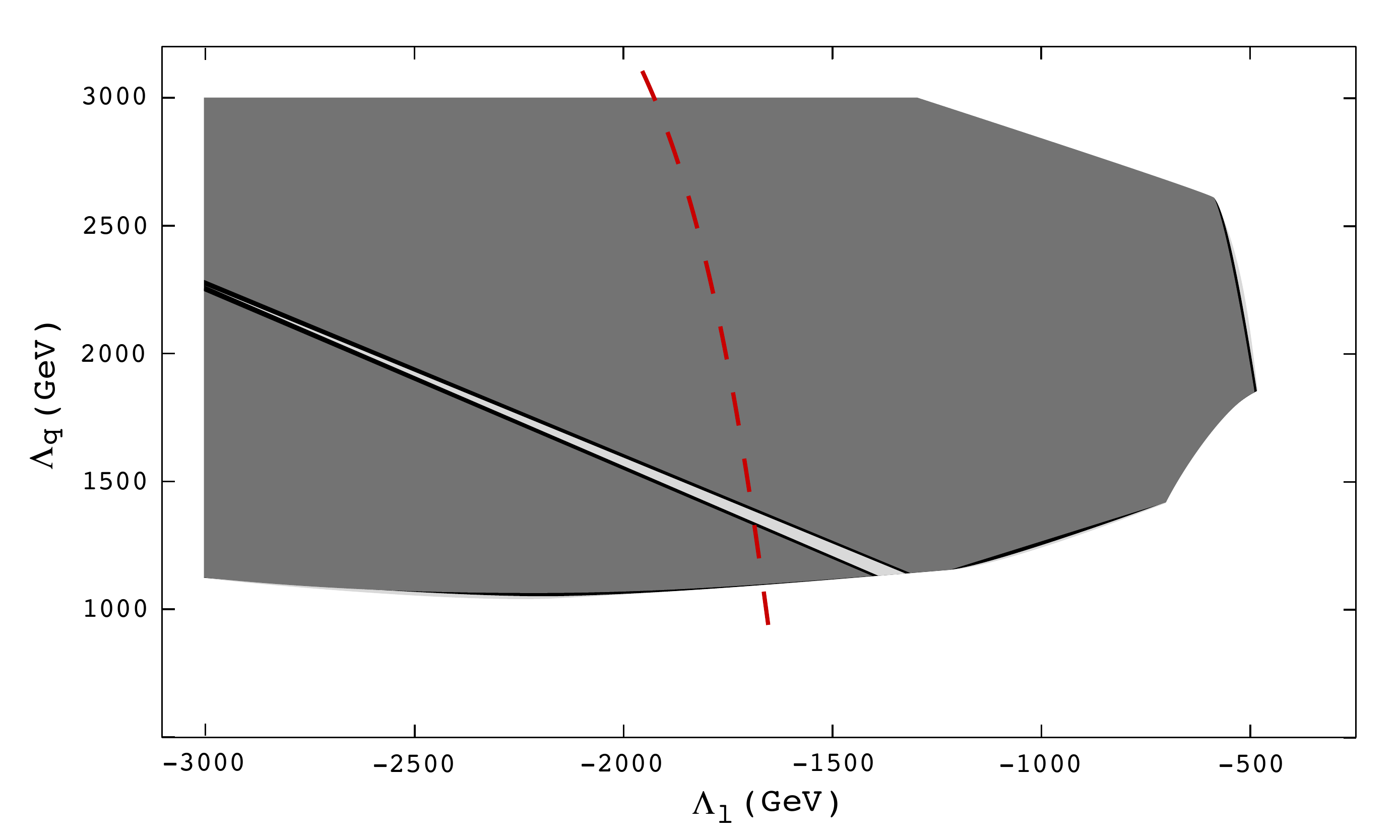} 
   \caption{Parameter space for neutralino dark matter in a two-parameter model of ``general'' sequestered gauge mediation with scalar sequestering as a function of the parameters $\Lambda_q$ and $\Lambda_l$. Adequate relic abundance arises when the bino becomes light. Here $\sqrt{F_S} = 10^{11}$ GeV, $\tan \beta = 10.0,$ $\mu = 1$ TeV, $B \mu \simeq \mu^2,$ and $ \tilde \Lambda_{q,l}^2 \simeq \Lambda_{q,l}^2/16 \pi^2.$ Color conventions are as in Fig. \ref{fig:min}, except here $m_h > 111.4$ GeV to the {\it left} of the dotted red line.}
   \label{fig:ggm0}
\end{figure}

\begin{figure}[t] 
   \centering
   \includegraphics[width=4in]{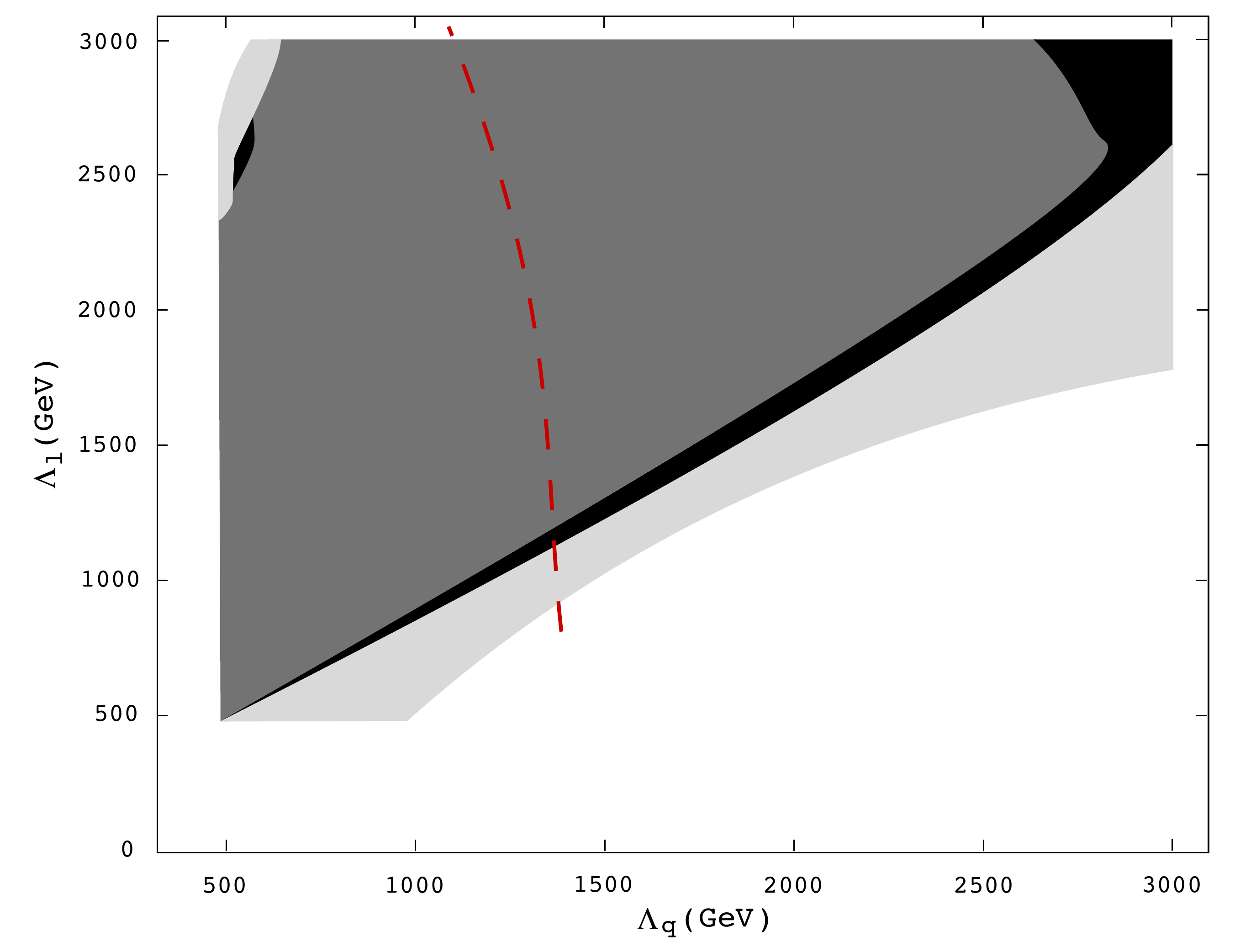} 
   \caption{Parameter space for neutralino dark matter in a two-parameter model of ``general'' sequestered gauge mediation as a function of the parameters $\Lambda_q$ and $\Lambda_l$. We assume some other mechanism for solving the $\mu$ problem. Here $\sqrt{F_S} = 10^{11}$ GeV, $\tan \beta = 10.0,$ $B \mu \simeq \mu^2,$ and $\Lambda_{q,l}^2 \simeq \tilde \Lambda_{q,l}^2.$ Color conventions are as in Fig. \ref{fig:min}.}
   \label{fig:ggm1}
\end{figure}

The result is a simple theory in which squark and slepton masses at the SUSY-breaking scale are suppressed relative to the gaugino masses, $m_{\tilde f} \simeq  \left( \frac{\Lambda_*}{M_*} \right)^{\frac{1}{2}\gamma_{S^\dagger S}- \gamma_S} M_a.$
In general, the operator giving rise to $B$ is suppressed relative to $\mu$ by a factor of  $\left( \frac{\Lambda_*}{M_*} \right)^{\frac{1}{2} \gamma_{S^\dagger S} - \gamma_S}$ as well;  any solution to the $\mu$ problem from strong coupling in the hidden sector suppresses high-scale squark and slepton masses by a similar amount relative to the gaugino masses. The parameter space for neutralino dark matter in two such theories is shown in Figs. \ref{fig:ggm0} and \ref{fig:ggm1}. 

An analogous three-parameter model may be obtained if the $5 + \bar 5$ of the messenger sector is replaced by a $10 + \bar{10}$ with corresponding superpotential
\beq
W = S \left(\lambda_Q \bar Q Q + \lambda_U \bar U U + \lambda_E \bar E E \right) + M (\bar Q Q + \bar U U + \bar E E)
\eeq
In this case, there predictably arise three gaugino mass parameters
\beq
\Lambda_I = \lambda_I \frac{ F}{M}  \left( \frac{\Lambda_*}{M_*} \right)^{\gamma_{S}}  \hspace{5mm} 
\eeq
where $I = Q, U, E.$ The gaugino masses are given by
\beq
M_1 &=& \frac{\alpha_1}{4 \pi} \left[ \frac{1}{3} \Lambda_Q + \frac{8}{3} \Lambda_U + 2 \Lambda_E \right] \\
M_2 &=&  \frac{\alpha_2}{4 \pi} 3 \Lambda_Q \nonumber \\
M_3 &=& \frac{\alpha_3}{4 \pi} \left( 2 \Lambda_Q + \Lambda_U \right) \nonumber
\eeq
Analogously, the squark and slepton parameters are 
\beq
\tilde \Lambda_I^2 \equiv \left| \frac{\lambda_I F}{M} \right|^{2} \left( \frac{\Lambda_*}{M_*} \right)^{\gamma_{S^\dagger S}}
\eeq
and the corresponding masses given by
\beq
m_{\tilde f}^2 = 2  \left[ C_1^{\tilde f} \left( \frac{\alpha_1}{4 \pi} \right)^2  \left(\frac{1}{3} \tilde \Lambda_Q^2 + \frac{8}{3} \tilde \Lambda_U^2 + 2 \tilde \Lambda_E^2 \ \right) + C_2^{\tilde f} \left( \frac{\alpha_2}{4 \pi} \right)^2 3 \tilde \Lambda_Q^2 \right. \\
\left. + C_3^{\tilde f} \left( \frac{\alpha_3}{4 \pi} \right)^2 \left(2 \tilde \Lambda_Q^2 + \tilde \Lambda_U^2 \right) \right] \nonumber.
\eeq
We may generate $\mu$ and $B \mu$ terms for the Higgs by coupling directly to the messengers via a superpotential
\beq
W \supset \lambda H_u \, 10 \, 10 + \bar \lambda H_d \, \bar{10} \, \bar{10}
\eeq
Here, e.g., $10 = (Q, U, E).$ Due to the direct coupling to messengers, the Higgses obtain $\mu$ and $B \mu$ at one loop; in general, the values of $\mu$ and $B \mu$ are different for the doublet and triplet Higgses. The $\mu$ term for the Higgs doublets takes the form
\beq
\mu = \frac{\lambda \bar \lambda}{16 \pi^2} f\left(\frac{\lambda_Q}{\lambda_U}\right) \frac{F}{M}  \left( \frac{\Lambda_*}{M_*} \right)^{\gamma_S}
\eeq
where $f(x) = \frac{x \log x^2}{1-x^2}.$ Similarly, $B \mu$ for the Higgs doublets takes the form 
\beq
B \mu = \frac{\lambda \bar \lambda}{16 \pi^2} f\left(\frac{\lambda_Q}{\lambda_U}\right) \frac{F^2}{M^2}  \left( \frac{\Lambda_*}{M_*} \right)^{\gamma_{S^\dagger S}}
\eeq
As before, the $\mu$ problem in this theory may be resolved provided $\gamma_{S^\dagger S} > 2 \gamma_S.$

Combining the models with both $5 + \bar{5}$ and $10 + \bar{10}$ messengers yields a theory with five of the parameters of General Gauge Mediation, non-gravitino LSP, and a strongly-coupled solution to the $\mu$ problem. In these weakly-coupled models of general gauge mediation and their extensions, the unified prediction for gaugino masses is lost, allowing a parametric separation of bino and wino masses at the electroweak scale. This may significantly expand the parameter space of neutralino dark matter, since a relatively larger wino mass causes left-handed slepton masses to run higher. A potential drawback of this class of theory is that there may exist a relative phase between $\Lambda_q, \Lambda_l,$ (or $\Lambda_Q, \Lambda_U, \Lambda_E$) leading to prohibitive EDMs for quarks and leptons at one loop. However, it is worth emphasizing that this concern is true only of the weakly-coupled theories, and may not be true of the strongly-coupled messenger sectors they are meant to parameterize.

\section{Discussion: Genera of Neutralino Dark Matter  \label{sec:dm}}

	The relic abundance of supersymmetric dark matter is influenced by myriad processes, arising from both the detailed physics of the sparticle spectrum and the particular physics of dark matter annihilation. Consequently, there are a variety of ways to achieve dark matter relic abundance compatible with WMAP observational bounds. Even within the relatively simplistic framework of mSUGRA, there are regions of parameter space in which qualitatively different mechanisms give rise to neutralino dark matter with adequate relic density. The very same qualitative features that produce a wide variety of neutralino dark matter in mSUGRA and its descendents may arise in gauge mediation if strongly coupled dynamics in the hidden sector allow for neutralino LSP. Consequently, the richness of detectable weak-scale dark matter candidates arising from gravity-mediated supersymmetry breaking may naturally extend to gauge mediation. 
	
	In the simplest single-messenger model of sequestered gauge mediation, the parameter space for acceptable neutralino dark matter is relatively constrained; adequate relic abundance is achieved only through coannihilations. With the added parametric freedom of  non-minimal gauge-mediated models such as those considered in Sec. \ref{sec:models}, it is possible to realize more varied genera of neutralino dark matter. As such, sequestered gauge mediation makes it possible to realize the full phenomenological appeal of mSUGRA and its relatives without troublesome flavor problems. Here we briefly consider some of the phenomenological range of dark matter allowed by sequestered gauge mediation.	
	
\subsection{Bino dark matter}

Much as in gravity-mediated scenarios, it is often the case that a bino-like neutralino is the LSP in much of the parameter space for non-minimal models of sequestered gauge mediation. In the so-called `bulk' region, where ancillary processes like coannihilation are insignificant and the principal decay occurs through t-channel sfermion exchange, the relic abundance of bino-like LSP is too high over a wide range of parameters. However, since $\langle \sigma v \rangle \sim m_\chi^2,$ the neutralino relic abundance may be made suitably small in this bulk region if the neutralino mass is sufficiently low. At such low masses, the bulk region is generally disfavored by collider data such as LEP sparticle mass limits and Higgs mass bound. These constraints may be evaded in theories without gaugino mass unification, in which case there are few limits on the mass of a light bino \cite{Dreiner:2009ic}. As a result, suitable low-mass bino dark matter in the `bulk' region may arise in models of general gauge mediation without gaugino mass unification, as seen in Fig. \ref{fig:ggm0}. 

There are, of course, ways to lower the bino relic abundance through more efficient annihilation channels. It is well known that the neutralino relic density may be significantly lowered by coannihilations in the presence of additional, nearly-degenerate particles \cite{Binetruy:1983jf, Griest:1990kh}. If additional particles are present whose masses are close to that of the LSP (i.e., the mass difference is of the same order as the freezout temperature $T_f$), then their annihilations may significantly lower the resultant dark matter relic abundance. In the case of sequestered gauge mediation, the stau and sneutrino are often nearly degenerate with the neutralino due to the smallness of scalar masses at the SUSY-breaking scale, and coannihilation of the stau and sneutrino at the borders of neutralino parameter space may significantly decrease the neutralino relic abundance. Obtaining the correct relic abundance requires a relatively small splitting, $(m_{\tilde \tau} - M_1)/M_1 \lesssim T_f / M_1 \sim 5 \%$. Indeed, in the minimal model of sequestered gauge mediation, coannihilation is significant throughout the narrow region of neutralino parameter space and crucial to obtaining adequate relic abundance. This may be seen vividly in Fig. \ref{fig:min}; suitable relic abundance arises only on the edges of parameter space for the neutralino LSP, where the lightest neutralino is nearly degenerate with the stau or sneutrino. In generalized models of sequestered gauge mediation, the stau may be significantly heavier than the neutralino, but coannihilations still play a significant role at the borders of neutralino parameter space (as seen, e.g., in the upper region of Fig.  \ref{fig:mm1}).

Bino-like neutralino relic abundance may also be significantly reduced by resonant annihilations. One such dominant process is annihilation through the Higgs pseudoscalar $A,$ which may mediate $s$-channel annihilation of the neutralino into fermions \cite{Baer:2002fv, Baer:2002ps}. These annihilations are efficient for $2 m_{\chi} \approx m_A,$ and may likewise expand significantly the region of neutralino relic abundance satisfying WMAP bounds. However, resonant light-Higgs exchange in sequestered gauge mediation is almost entirely ruled out by LEP constraints.

\subsection{Higgsino dark matter}

\begin{figure}[t] 
   \centering
   (a)
   \includegraphics[height=2.5in]{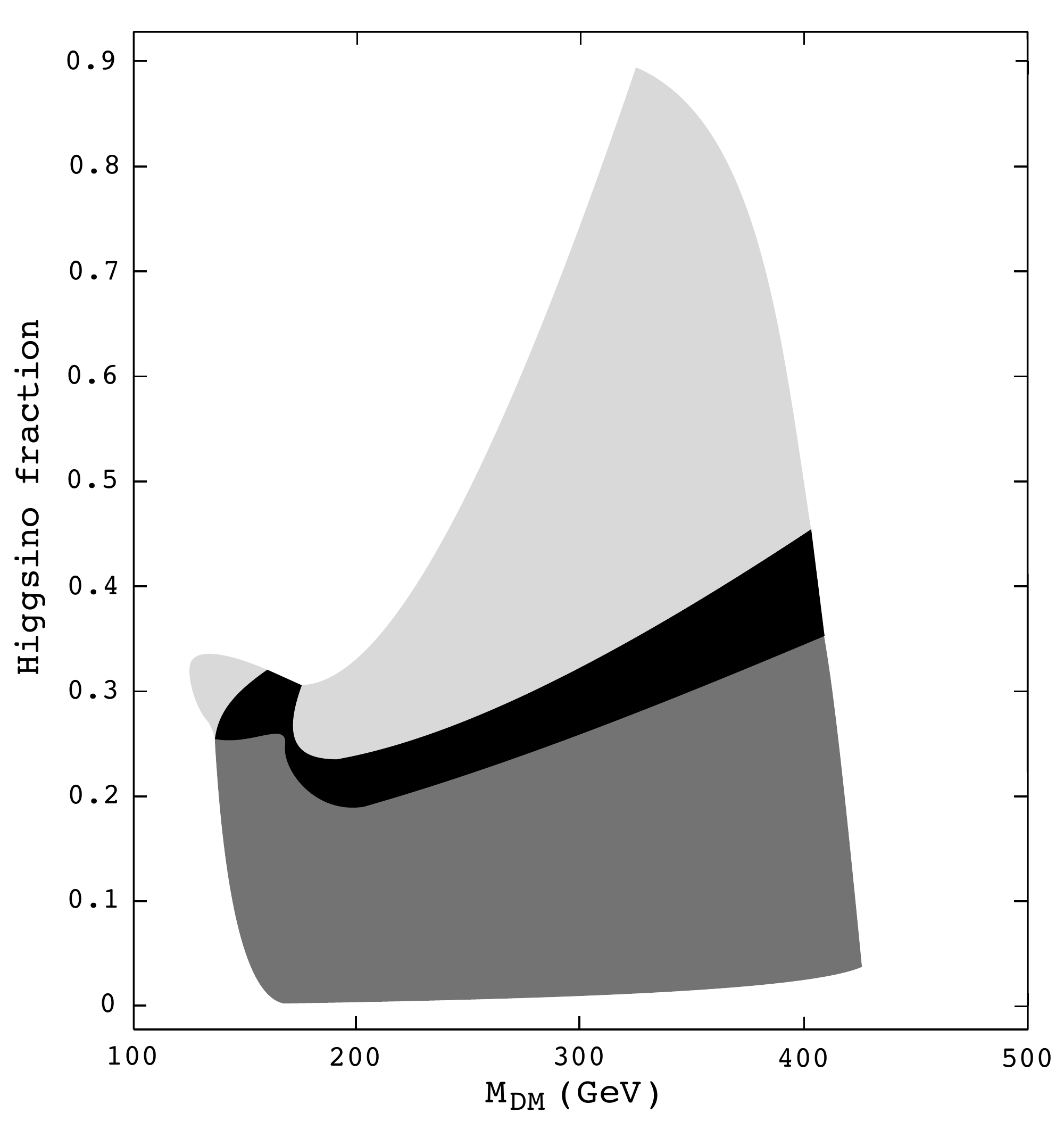} 
   (b) 
      \includegraphics[height=2.5in]{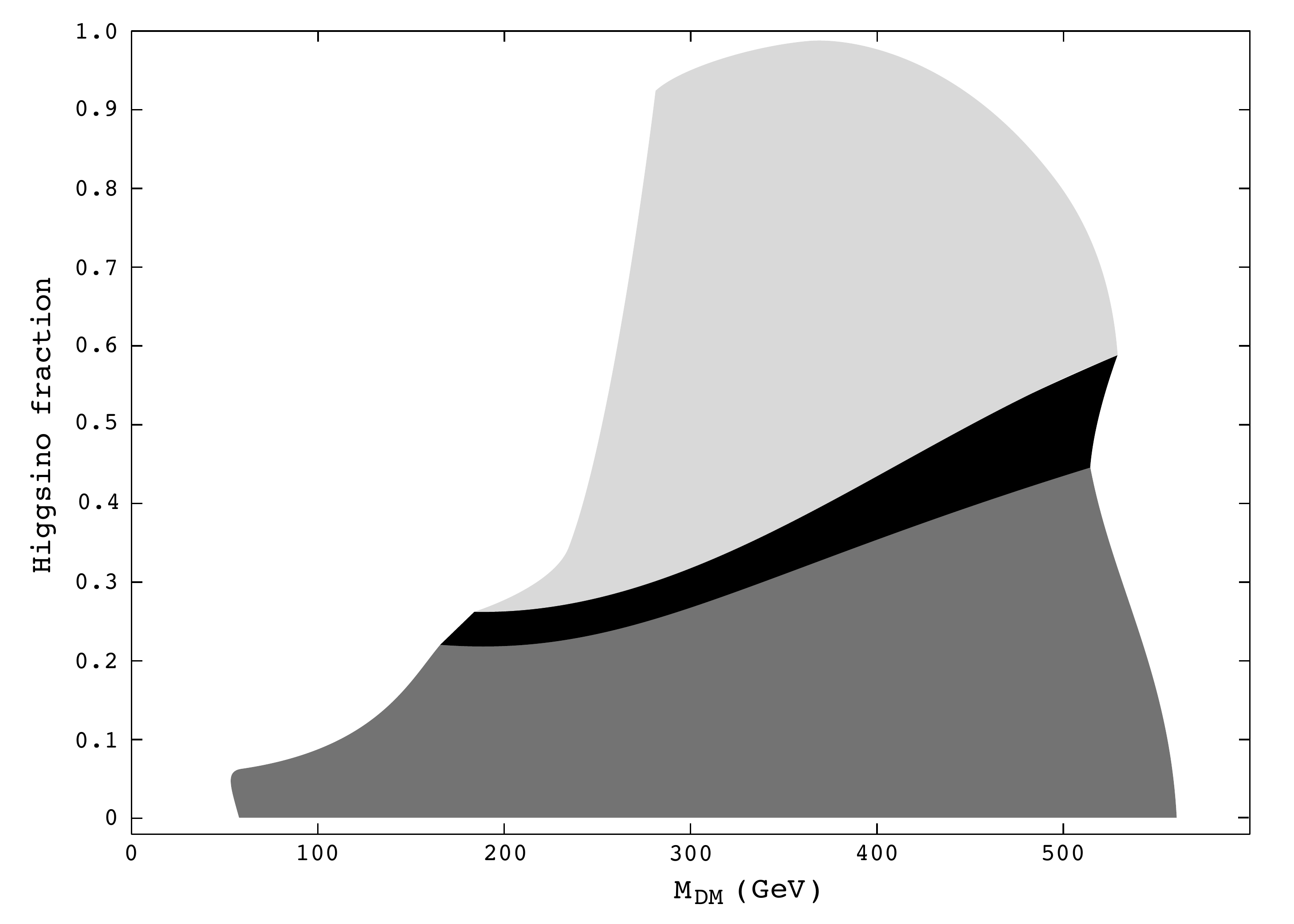} 
   \caption{Neutralino dark matter as a function of dark matter mass $m_{DM}$ and higgsino fraction $\sqrt{|Z_{13}|^2 + |Z_{14}|^2}$ for (a) the parameter space of Fig. \ref{fig:nbm} and (b) the parameter space of Fig. \ref{fig:mm1} (wino fraction is negligible in both cases). Color conventions are as in Fig. \ref{fig:min}. Relic abundance for mostly-bino neutralino is too high; that for mostly-higgsino neutralino is too low; suitable relic abundance is obtained for a well-tempered mix of bino and higgsino.}
   \label{fig:hdm}
\end{figure}

Of course, it is not always the case that the lightest neutralino is bino-like. In theories with gravity-mediated supersymmetry breaking, the higgsino is an infrequent contender assuming universality of SUSY-breaking soft parameters. However, for large values of the scalar mass parameter $m_0,$ there may arise ``focus points" in the RG running of the Higgs mass $m_{H_u}^2,$ such that its value at the electroweak scale is relatively insensitive to its original value in the ultraviolet \cite{Feng:1999mn, Feng:1999zg}. Although the scalar mass parameter $m_0$ is large in this scenario, the value of $|\mu|$ remains small. Consequently, the higgsino fraction of the lightest neutralino is enhanced. Moreover,  small values of $\mu$ are naturally allowed if universality of soft masses is no longer imposed, leading to higgsino LSP without recourse to specific features of RG flow. Whatever the origin, a significant higgsino component of the neutralino LSP yields more efficient annihilations through such processes as the $t$ channel decay $\chi_0 \chi_0 \rightarrow \chi^+ \rightarrow W^+ W^- ,$ leading to a reduced relic abundance. The near-degeneracy between neutral and charged higgsinos makes coannihilations inevitable, and chargino mass limits constrain lighter versions of higgsino dark matter . 

Although the RG running of MSSM fields in sequestered gauge mediation is significantly different from that of mSUGRA, it is often the case that large scalar masses and small values of $\mu$ arise naturally due to differing degrees of sequestering among the various soft masses. Such behavior was observed even in the simplest single-messenger case \cite{Craig:2008vs}, albeit in a relatively constrained region of parameter space yielding insufficient relic abundance. As in the case of mSUGRA, the efficiency of annihilations into gauge bosons, as well as the inevitable coannihilations between higgsino and chargino, make the relic abundance too low in pure-higgsino regions of LSP parameter space. This is the origin of the too-low relic abundance for $\mu \ll M_0$ seen in Figs. \ref{fig:nbm} and \ref{fig:mm1}; the direct relation between higgsino fraction and relic abundance for both theories is shown clearly in Fig. \ref{fig:hdm}.

\subsection{Wino dark matter}

\begin{figure}[t] 
   \centering
   \includegraphics[height=2.5in]{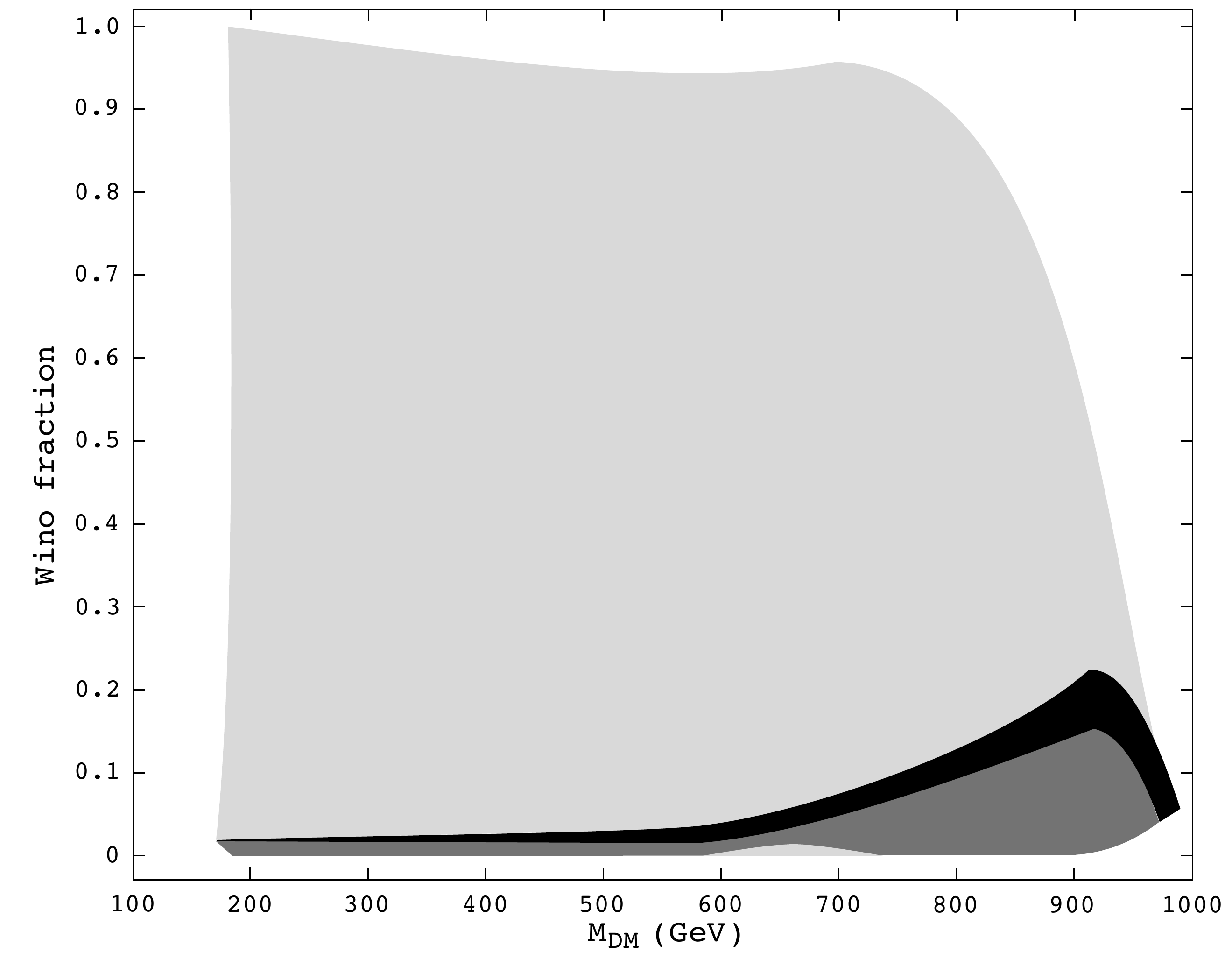} 
   \caption{Neutralino dark matter as a function of dark matter mass $m_{DM}$ and wino fraction $|Z_{12}|^2$ for the parameter space of Fig. \ref{fig:ggm1} (higgsino fraction is negligible). Color conventions are as in Fig. \ref{fig:min}. Relic abundance for mostly-bino neutralino is too high; that for mostly-wino neutralino is too low; suitable relic abundance is obtained for a well-tempered mix of bino and wino.}
   \label{fig:ggm1dm}
\end{figure} 

A third -- and more exotic -- possibility is for the LSP to be predominantly wino. While impossible to attain in minimal sequestered gauge mediation assuming gaugino mass unification, the wino may be the LSP in more general models such as the weakly-coupled GGM theories of Sec. \ref{sec:models}. In these situations the wino LSP tends to arise only when there is no scalar sequestering (i.e., when $2 \gamma_S \gtrsim \gamma_{S^\dag S}$); for such theories {\it with} scalar sequestering, the sneutrino is invariably lighter than the wino-like neutralino. In any event, the efficiency of wino annihilations is even more extreme than that of higgsinos, and relic abundance for mostly-wino dark matter is far too low for all but the most unnaturally heavy of candidates (see, e.g., Figs. \ref{fig:ggm1} and \ref{fig:ggm1dm}).

\subsection{`Well-tempered' dark matter}

As noted in \cite{ArkaniHamed:2006mb}, the observed dark matter relic abundance is most readily obtained by a `well-tempered' mixture of bino-higgsino or bino-wino LSP. Although such scenarios appear somewhat tuned since they require a precise relationship among parameters (either $|M_1| \simeq |\mu|$ or $|M_1| \simeq |M_2|$), they do seem to be among the few remaining possibilities not relying upon resonant annihilations to obtain acceptable relic abundance. Both types of well-tempered neutralino may be realized in models of sequestered gauge mediation. In theories preserving gaugino mass unification, such as those with messengers in complete $SU(5)$ multiplets, only the bino-higgsino neutralino is possible; in theories with messengers in partial multiplets, such as weakly-coupled models of general gauge mediation, it is possible to obtain either bino-higgsino or bino-wino neutralino LSP.

Typically, this `well-tempered' neutralino is the most likely dark matter candidate in more general theories of sequestered gauge mediation. Since these theories exhibit a neutralino LSP over wide ranges of parameter space, the conventional problems of mostly-bino or mostly-higgsino dark matter lead to too-high or too-low thermal abundances, respectively. While tuned phenomena such as coannihilations and resonant annihilations operate in small regions of parameter space, it is most often the case that suitable relic abundance results from neutralinos that are an admixture of bino and higgsino. The suitability of `well-tempered' neutralinos is shown most vividly for mixed bino-higgsino dark matter in Fig. \ref{fig:hdm} and for mixed bino-wino dark matter in Fig. \ref{fig:ggm1dm}.

\section{Conclusion \label{sec:conc}}

Strong dynamics in the hidden sector may significantly alter the sparticle spectrum of the MSSM. In the case of gauge-mediated supersymmetry breaking these strong dynamics may raise the gravitino mass relative to gaugino masses, allowing for weak-scale neutralino dark matter without prohibitive flavor violation. 

While the parameter space for neutralino dark matter is tightly constrained in the simplest single-messenger theory of sequestered gauge mediation, more elaborate messenger sectors exhibit a far greater degree of parametric freedom. In particular, such non-minimal messenger sectors alter the naive relations between gaugino, scalar, and Higgs-sector masses. As a result, strong dynamics in the hidden sector may simultaneously provide compelling neutralino dark matter candidates and a solution to the $\mu / B \mu$ problem without prohibitive fine-tuning. 

In this paper, we have explored various generalizations of gauge mediation with strongly coupled hidden sectors. Such theories do not require exceptionally baroque dynamics or unusual relations between supersymmetry-breaking soft masses. Depending on the details of the messenger sector, the full richness of supersymmetric dark matter phenomenology familiar from mSUGRA may be realized without flavor violation or a $\mu / B \mu$ problem. Neutralino dark matter with the adequate relic abundance may arise via coannihilations, resonant processes, or suitable mixing among gauginos and higgsinos. These dark matter candidates are consistent with viable electroweak symmetry breaking and existing experimental constraints. Both the neutralino dark matter candidates and the distinctive sparticle spectra arising from theories of sequestered gauge mediation will soon be within reach of direct detection experiments and production at the LHC. 

Our focus has been largely limited to parameterizing general models of sequestered gauge mediation and their neutralino dark matter candidates. It would be extremely useful to conduct a more complete study of these various models, their typical sparticle spectra, and expected signatures at the LHC. It would also be worthwhile to look far beyond the MSSM. Although our analysis has centered on the MSSM and its conventional dark matter candidates, the phenomenology of sequestering explored in this paper is rather general. Similar physics may play a role in more exotic dark sectors, a possibility that merits further investigation.

\acknowledgments
We would like to thank Savas Dimopoulos, Michael Dine, Shamit Kachru, Hyung Do Kim, John Mason, Michael Peskin, Eva Silverstein, and David Poland for helpful discussions. DG is supported by NSERC, the Mellam Family Foundation, the DOE under contract DE-AC03-76SF00515 and the NSF under contract PHY-9870115. NJC is supported by the NSF GRFP, the NSF under contract PHY-9870115, and the Stanford Institute for Theoretical Physics.

\bibliography{gmdm2}
\end{document}